\documentclass[twocolumn,amsmath,amssymb]{snp}
\usepackage{graphicx}
\usepackage{dcolumn}
\usepackage{bm}
\begin{document}

\title{{\Large Role of nuclear radii  in the exotic cluster-decay }}

\author{\large Narinder K. Dhiman}
\email{narinder.dhiman@gmail.com}
\affiliation{Govt. Sr. Sec. School, Summer Hill, \\
 Shimla -171005, INDIA}
\maketitle
\section*{Introduction}
Recently, a renewed interest has emerged in nuclear physics
research. This includes low energy fusion process, intermediate
energy phenomena as well as cluster-decay and/or formation of super
heavy nuclei~\cite{Pra}. In the last one decade, several theoretical
models have been employed in the literature to estimate the
half-life times of various exotic cluster decays of radioactive
nuclei.  In these models, one needs complete knowledge of the
interaction potential as well as of nuclear radii. Our aim here is
to study the effect of various nuclear radii forms available in the
literature~\cite{r1,r2,r3,r4,r5,r6,r7} on the cluster decay process.
\section*{The Model}
In the spirit of proximity force theorem~\cite{nkt}, the nuclear
potential $V_{N}(R)$ of the two spherical nuclei, with radii C$_1$
and C$_2$ and whose centers are separated by a distance
$R=s+C_1+C_2$ is given by
\begin{equation}
V_{N}\left(R \right)  = 2\pi \overline{R} \phi (s), \label{eq:1}
\end{equation}
where
\begin{equation}
\phi (s) =\int \left\{H(\rho)- \left[H_{1}(\rho_{1}) +
H_{2}(\rho_{2}) \right] \right\}dZ, \label{eq:2}
\end{equation}
and
\begin{equation}
\overline{R} =\frac{C_1 C_2}{C_1+C_2}, \label{eq:3}
\end{equation}
with S\"ussmann central radii C$_i$ given in terms of equivalent
spherical radii R$_i$ as:
\begin{equation}
C_i =R_i -b{R_i}^{-1}. \label{eq:4}
\end{equation}
Here surface diffuseness parameter $b=1$ fm and nuclear radii R$_i$
taken as given by various authors~\cite{r1,r2,r3,r4,r5,r6,r7}. These
different nuclear radii are labeled as R$_{Prox77}$~\cite{r1},
R$_{AW}$~\cite{r2}, R$_{Prox00}$~\cite{r3}, R$_{Royer}$~\cite{r4},
R$_{Ngo}$~\cite{r5}, R$_{Bass}$~\cite{r6} and R$_{CW}$~\cite{r7}.

For the cluster decay calculations, we use the Preformed Cluster
Model (PCM) based on the well known quantum mechanical fragmentation
theory~\cite{Pra}. The decay constant $\lambda$ or decay half-life
$T_{1/2}$, is defined as:
\begin{equation}
\lambda =\frac{\ln 2}{T_{1/2}}= P_{0}\nu _{0}P,
 \label{eq:5}
\end{equation}
where $P_0$, $P$ and  $\nu_0$  refers to the preformation
probability, the penetrability and assault frequency, respectively.

For decoupled hamiltonian, the Schr\"odinger equation in
$\eta$-co-ordinates can be written as:
\begin{equation}
[-\frac{\hbar ^{2}}{2\sqrt{B_{\eta\eta }}}\frac{\partial }{\partial
\eta }\frac{1}{\sqrt{B_{\eta \eta }}}\frac{\partial }{\partial \eta
}+V_R(\eta)]\psi(\eta)=E \psi(\eta). \label{eq:6}
\end{equation}

\section*{Results and Discussion}
We present here the cluster decay calculations of $^{56}$Ni, when
formed in heavy-ion collisions. Since $^{56}$Ni has negative
$Q$-value (or $Q_{out}$) and is stable against both fission and
cluster decay processes. Since it has a negative $Q_{out}$ having
different values for various exit channels and hence would decay
only when it is produced with sufficient compound nucleus excitation
energy $E^{\ast}_{CN}~(=E_{cm} + Q_{in})$, to compensate for
negative $Q_{out}$, their total kinetic energy ($TKE$) and the total
excitation energy ($TXE$) in the exit channel as:
\begin{equation}
E^{\ast}_{CN} = \mid Q_{out} \mid + TKE + TXE,
 \label{eq:7}
\end{equation}
\begin{figure}
\includegraphics[scale=0.31]{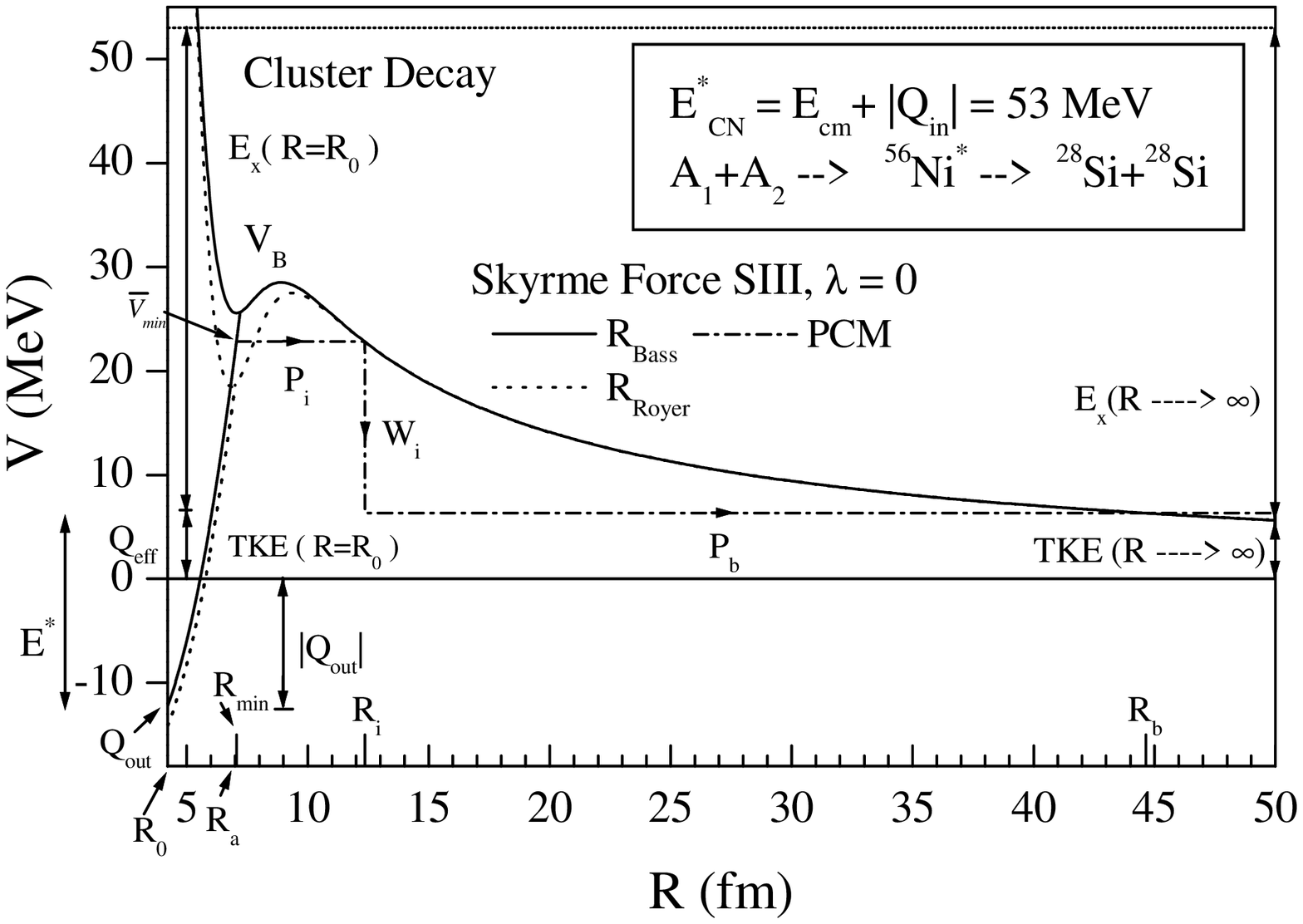}
\caption{\label{fig1} The scattering potential $V(R)$ (MeV) for
cluster decay of $^{56}$Ni$^{\ast}$ into $^{28}$Si + $^{28}$Si
channel for different nuclear radii. The   distribution of compound
nucleus excitation energy E$_{CN}^{*}$ at both the initial
($R=R_{0}$) and asymptotic ($R \to  \infty$) stages and $Q$-values
are shown. The decay path for PCM is also displayed.}
\end{figure}
here fragments are considered to be spherical  so the deformation
energy $E_d$ is neglected  (see Fig.~\ref{fig1}).  Here $Q_{in}$
adds to the entrance channel kinetic energy $E_{cm}$ of the incoming
nuclei in their ground states.

Fig.~\ref{fig1} shows the characteristic scattering potential for
the cluster decay of $^{56}$Ni$^{\ast}$ into $^{28}$Si + $^{28}$Si
channel for two different forms of nuclear radii (R$_{Bass}$ \&
R$_{Royer}$). In the exit channel, for the compound nucleus to
decay, the compound nucleus excitation energy $E_{CN}^{\ast}$ goes
in compensating the negative $Q_{out}$, the total excitation energy
$TXE$ and total kinetic energy $TKE$ of the two outgoing fragments.
The $TKE$ plays the role of effective Q-value ($Q_{eff}$)  in the
cluster decay process. We plot the penetration path for PCM using
Skyrme force SIII (without surface correction factor, $\lambda =0$)
with nuclear radius R$_{Bass}$. We begin the penetration path at
$R_a = R_{min}$ with potential at this $R_a$-value as $V(R_a =
R_{min})= \overline{V}_{min}$ and ends at $R = R_b$, corresponding
to $V(R=R_b) = Q_{eff}$.  The $Q_{eff}$ values are taken from
ref.~\cite{Pra}.
\begin{figure}
\includegraphics[scale=0.31]{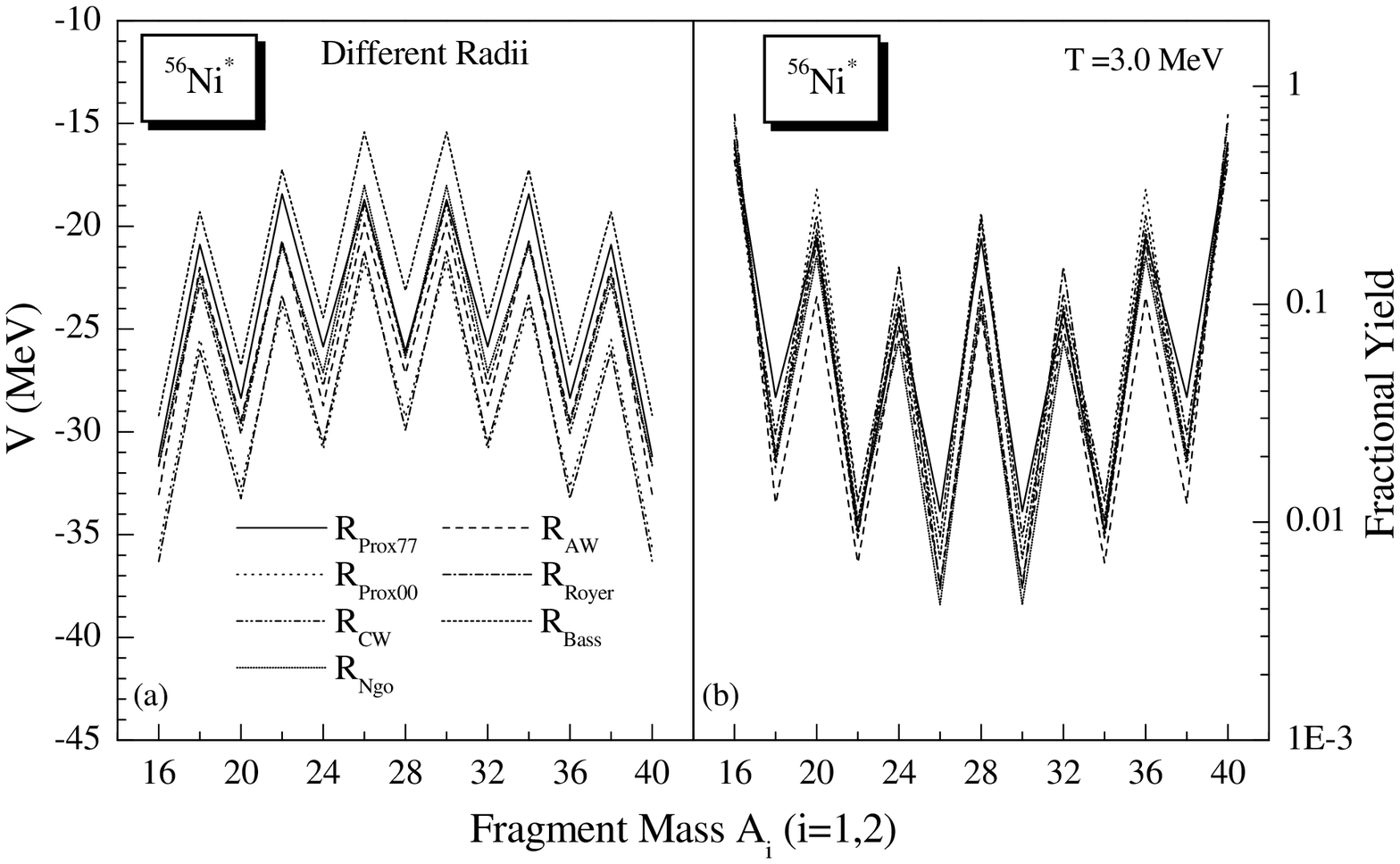}
\caption{\label{fig2} (a) The fragmentation potential $V(\eta)$ and
(b) calculated fractional mass distribution yield with different
 nuclear radii at $T$ = 3.0 MeV.}
\end{figure}

Fig.~\ref{fig2}(a) and (b) shows the fragmentation potential
$V(\eta)$ and fractional mass distribution yield at $R = R_{min}$
with $V(R_{min})= \overline{V}_{min}$. The fractional yields are
calculated within PCM at $T$ = 3.0 MeV for $^{56}$Ni$^{\ast}$ using
various forms of nuclear radii. From figure, we observe that
different radii gives approximately similar behavior, however, small
changes in the fractional mass distribution yields are also
observed. The fine structure is not at all disturbed for different
radius values.

We have also calculated the half-lives (or decay constants) of
$^{56}$Ni$^{\ast}$ within PCM for clusters $\ge ^{16}$O. The
variation in the cluster decay half-lives studied with respect to
radius formula due to Royer (R$_{Royer}$) shows that the half-lives
lie within $\pm$7\%, excluding Bass radius formula where these lie
within $\pm$10\%~\cite{nkd}.

\end{document}